\documentclass[final,5p,times,twocolumn]{elsarticle}

\usepackage{xcolor}
\usepackage{amsmath}
\usepackage{amssymb}
\let\amssBox\Box
\usepackage{wasysym}
\let\wasyPentagon\pentagon
\usepackage{bm}
\usepackage{dcolumn}
\usepackage{textcomp}
\usepackage[dvipsnames]{xcolor}
\usepackage{color}
\usepackage{multirow}
\usepackage{booktabs}
\usepackage{braket}
\usepackage{graphicx}
\usepackage{upgreek}
\usepackage{soul}
\usepackage{physics}
\usepackage{url}
\usepackage{enumitem}
\biboptions{sort&compress}
\definecolor{C0}{HTML}{1f77b4}
\definecolor{C1}{HTML}{ff7f0e}
\definecolor{C2}{HTML}{2ca02c}
\definecolor{C3}{HTML}{d62728}
\definecolor{C4}{HTML}{9467bd}
\definecolor{C5}{HTML}{8c564b}
\definecolor{C6}{HTML}{e377c2}
\definecolor{C7}{HTML}{7f7f7f}
\definecolor{C8}{HTML}{bcbd22}
\definecolor{C9}{HTML}{17becf}
\definecolor{DeepPink}{HTML}{ff1493}
\usepackage[pdfstartview=FitH, colorlinks=true]{hyperref}
\usepackage[table]{xcolor}
\usepackage{orcidlink}

\journal{Physics Letters B}

\begin{document}
\begin{frontmatter}

\title{Complex-energy eigenvector continuation for nuclear many-body broad resonances}

\author[ad1]{R. Z. Hu\,\orcidlink{0009-0002-8797-6622}}
\author[ad2,ad3,ad4]{N. Michel\,\orcidlink{0000-0003-4214-8620}}
\author[ad1]{Z. C. Xu\,\orcidlink{0000-0001-5418-2717}}
\author[ad2,ad3]{J. G. Li\,\orcidlink{0000-0001-6002-7705}}
\author[ad1,ad2,ad3]{F. R. Xu\,\orcidlink{0000-0001-6699-0965}\corref{cor1}}

\address[ad1]{School of Physics, and State Key Laboratory of Nuclear Physics and Technology, Peking University, Beijing 100871, China}
\address[ad2]{Institute of Modern Physics, Chinese Academy of Sciences, Lanzhou 730000, China}
\address[ad3]{Southern Center for Nuclear-Science Theory, Institute of Modern Physics, Chinese Academy of Sciences, Huizhou 516000, China}
\address[ad4]{School of Nuclear Science and Technology, University of Chinese Academy of Sciences, Beijing 100049, China}
\cortext[cor1]{frxu@pku.edu.cn}

\begin{abstract}
Broad resonances are a unique phenomenon in nuclear many-body systems. Theoretical studies usually involve the continuum degree of freedom, which drastically increases the model space of calculations, and may lead to non-convergence or instability of computations. In this paper, we present the extension of the eigenvector continuation (EC) method to the complex-energy space to treat the broad resonances of open quantum systems of nuclei. EC provides an efficient method to predict the solution of a large-space many-body problem within a small subspace. Using only a few bound and narrow resonance solutions as input in EC, we can obtain the solution of a broad resonance. We have applied the complex-energy EC to the broad resonances of $^4$H, four-neutron $^4n$, $^6$He and $^7$He systems.
\end{abstract}


\end{frontmatter}

\section{Introduction}

While resonances are commonly found in nuclear many-body systems, some of them exhibit broad decay widths.
Nuclear broad resonances show strong coupling to the external environment, and undergo particle emissions within a very short time, thus belong to the category of so-called open quantum systems~\cite{Michel_2009}.
Theoretical studies of broad resonances are challenging current nuclear models~\cite{alphaexp2023,nicolasPRL2023,9Nexp2023}. Nuclear resonance which eventually decays through particle emission should be a time-dependent quantum mechanics problem, but it can be approximated by the time-independent quasi-stationary Schr{\"o}dinger equation by introducing the concept of complex energy~\cite{Gamow1928,PhysRev.56.750,BERGGREN1968265}. The complex-energy Berggren basis which contains bound, resonance and continuum single-particle states provides a powerful basis for nuclear many-body calculations with the continuum coupling incorporated at the basis level. The Gamow shell model (GSM) using the Berggren basis has shown great power in describing nuclear many-body resonances~\cite{PhysRevLett.89.042501,PhysRevLett.89.042502,SUN2017227,HU2020135206,MA2020135257,ZHANG2022136958}.

Broad resonances contain significant continuum ingredients. The inclusion of continuum partial waves results in a dramatic increase in the model space, which can lead to non-convergence or instability in computations. To find the physical solution of the broad resonance, one may add an external potential~\cite{PhysRevLett.90.252501,PhysRevLett.118.232501,PhysRevC.100.054313} or scale the interaction by multiplying a scaling factor~\cite{PhysRevLett.119.032501} so that one obtains an artificial bound or narrow resonance state for which the calculation is converged and stable. Then, one gradually reduces the external potential or scaling factor, and tracks the solution. But, one would still not be able to obtain the real solution of the broad resonance, and numerical extrapolation has to be used. It was commented that there are no promising extrapolation methods for calculations based on the complex-energy Berggren basis~\cite{KRASNOPOLSKY1978251,PhysRevLett.123.069201,PhysRevLett.123.069202}.

The recently developed eigenvector continuation (EC)~\cite{Frame2018} has shown excellent ability to solve a large-space many-body problem in a small subspace~\cite{Frame2018,PhysRevLett.123.252501,PhysRevC.101.041302,KONIG2020135814,Yapa2023}. The EC method has also been extended to unbound resonances of two- and three-body systems using schematic potentials~\cite{Yapa2023,Yapa2025}. In the present paper, we study the complex-energy EC using realistic interactions, focusing on broad resonances of many-body nuclear systems.

\section{Formalism}
The GSM calculation is the starting point of our EC process. The GSM uses the complex-energy Berggren basis~\cite{BERGGREN1968265}, thus incorporates the continuum effect at the basis level. For $^{6,7}$He we choose $^4$He as the core of the GSM calculations, while the no-core GSM (NCGSM) is used for $^4$H and four-neutron ($^4n$) resonances. The completeness relation of complex-energy (complex-momentum) Berggren basis states 
\begin{equation}
    \sum_{i} \vert u_{c}(k_i)\rangle \langle\tilde{u}_{c}(k_i)\vert
    + \int_{\mathcal{L}^{+}_{c}} d k \vert u_{c}(k)\rangle\langle \tilde{u}_{c}(k) \vert = \hat{1}_c
\end{equation}
is satisfied for each partial wave labeled by $c=(l,j)$, where $\vert u_{c}(k_i)\rangle$ are the radial wave functions of resonant states (including bound states and resonance poles), while $\vert u_{c}(k)\rangle$ are the  scattering states along a contour $\mathcal{L}^+_c$ in the fourth quadrant of the complex-momentum plane. The form of the contour is unimportant as long as the poles $\{k_i\}$ are all embedded between the real axis and the contour, guaranteed by the Cauchy's integral theorem~\cite{Michel2021}.

NCGSM is usually used for the resonances of $A\leq 5$ light nuclei~\cite{PhysRevC.88.044318,BARRETT2013131}, which is a natural extension of the no-core shell model in the complex-energy plane by replacing the harmonic oscillator basis with the Berggren basis.
The complex Hamiltonian matrix is not Hermitian but complex-symmetric, which has complex eigenvalues 
\begin{equation}
 E=E_R-i\Gamma/2   
\end{equation}
with $E_R$ and $-\Gamma/2$ being the real and imaginary parts of the eigen energy. $\Gamma$ describes the resonance width of the state, giving a particle emission half-life of $T_{1/2}=2\hbar \ln 2/\Gamma$.

To extend EC to nuclear resonances within the Berggren basis, we start with a scaled many-body Hamiltonian,
\begin{equation}
H(\alpha)=H_0 +\alpha H_1,
\label{H(alpha)}
\end{equation}
where $\alpha \geq 1$ is the scaling factor. In the NCGSM calculation, $H_0$ takes the intrinsic kinetic energy of the $A$-body nuclear system as
\begin{equation}
    H_0 = \frac{1}{A} \sum_{i<j}^A \frac{(\boldsymbol{p}_i-\boldsymbol{p}_j)^2}{2 m},
\end{equation}
where $\boldsymbol{p}_i$ is the nucleon momentum, and $m$ is the nucleon mass. Then, $H_1$ is the total potential of nucleon-nucleon (NN) interactions as
\begin{equation}
    H_1 = \sum_{i<j}^A V_{ij}^{\mathrm{NN}},
\end{equation}
where $V_{ij}^{\mathrm{NN}}$ is the NN interaction for which we use a realistic interaction. $\alpha=1$ gives the real Hamiltonian of the $A$-body nuclear system.

In the GSM with a core, to reduce the effect from the center-of-mass (COM) motion, the Hamiltonian is written in the cluster-orbital shell model (COSM) coordinates in which the positions of valence particles are relative to the COM of the core~\cite{PhysRevC.38.410,PhysRevC.96.054316},
\begin{equation}
    H=\sum_{i=1}^{A_v}\left[\frac{\boldsymbol{p}_i^2}{2 \mu_i}+U_{\mathrm{c}}(r_i)\right]
    + \sum_{i<j}^{A_v}\left[V_{ij}+\frac{\boldsymbol{p}_i \cdot \boldsymbol{p}_j}{M_{\mathrm{c}}}\right],
\label{H_COSM}
\end{equation}
where $A_\nu$ is the number of valence nucleons, and $\mu$ and $M_{\mathrm{c}}$ stand for the reduced mass of the nucleon and the mass of the core, respectively. $U_{\mathrm{c}}$ is the single-particle potential produced by the core~\cite{PhysRevC.96.054316}. $V_{ij}$ is the two-body interaction between valence nucleons. In the COSM coordinates, the translation invariance is preserved. However, the transformation of the realistic interaction to the COSM system has not been available~\cite{Michel_2009}. Therefore, in the GSM calculation with a core, we use the Furutani-Horiuchi-Tamagaki (FHT) potential for the interaction between valence nucleons outside the core~\cite{PhysRevC.96.054316}. Then, in the EC calculation based on the GSM with a core, with Hamiltonian~(\ref{H_COSM}), we demand
\begin{equation}
    H_0 =\sum_{i=1}^{A_v}\left[\frac{\boldsymbol{p}_i^2}{2 \mu_i}+U_{\mathrm{c}}(r_i)\right]
    + \sum_{i<j}^{A_v}\frac{\boldsymbol{p}_i \cdot \boldsymbol{p}_j}{M_{\mathrm{c}}},
\end{equation}
and
\begin{equation}
    H_1 =\sum_{i<j}^{A_v}V_{ij}.
\end{equation}

As starting points, we choose several $\alpha>1$ scaling factors with which the NCGSM or GSM with a core can yield converged numerical solutions for bound and narrow resonance states. These points $\{\alpha_i\}$ and obtained eigenvectors $\{\vert \psi_i \rangle\}$ are often referred to as ``training points'' and ``training vectors'' (or ``EC snapshots''), respectively. Then, for the target point $\alpha_*$ (i.e., $\alpha_*=1$ in the present work), we project the target Hamiltonian to the subspace spanned by the EC snapshots, resulting in a generalized eigenvalue problem~\cite{Frame2018,KONIG2020135814}
\begin{equation}
    \tilde{H}(\alpha_*)\big\vert\tilde{\psi}(\alpha_*)\big\rangle
    = \tilde{E}(\alpha_*) \tilde{N} \big\vert\tilde{\psi}(\alpha_*)\big\rangle,
\end{equation}
with the projected Hamiltonian and norm matrices being defined as
\begin{align}
     & \tilde{H}_{i j}(\alpha_*) = \langle\psi_i \vert H(\alpha_*) \vert \psi_j \rangle, \\
     & \tilde{N}_{i j} = \langle\psi_i \vert \psi_j \rangle=\int d p \, \psi_i(p) \psi_j(p)\label{norm},
\end{align}
where $\psi_i$ is the eigenvector of the target state at the training point of $\alpha_i>1$. The eigenvectors come from different Hamiltonians scaled with different $\alpha$ values. They are neither orthogonal nor normalized.

The scaled Hamiltonian has a linear dependence on the control parameter $\alpha$. It has been commented that a linear form well fits the EC process~\cite{Frame2018}. In the EC approximation, eigenvalues and eigenvectors out of reach can be obtained by projection via subspace learning~\cite{Frame2018}. This would provide a promising approach to solve a broad resonance problem in which the exact numerical calculation using nuclear models (e.g., GSM) may not converge or be unstable, whereas the computations of bound and narrow resonance states in subspace should be more stable. By choosing several proper training points $\alpha_i$, we obtain the corresponding training eigenvalues and eigenvectors of the subspace, which can be bound or narrow resonances. Then, we use the EC to project the broad resonance at the target $\alpha_* $ which is out of the training subspace.

\section{Calculations and discussions}
In the present work, we test EC calculations for the resonances of $^4$H, ${}^{4}n$, $^6$He and $^7$He. In $^4$H and ${}^{4}n$ calculations,
EC snapshots are provided by NCGSM calculations in the $spdfg$ space with the neutron $\nu$\{$s_{1/2}$, $p_{3/2}$\} using the Berggren basis and remaining higher partial waves using the harmonic oscillator (HO) basis. The proton in $^4$H is well bound, therefore we use the $spdfg$ HO basis for the proton. Such choice of the model space is reasonable and efficient~\cite{PhysRevC.100.054313,PhysRevC.104.024319}. In $^{6,7}$He calculations, EC snapshots are provided by GSM with the $^4$He core in the $spdf$ valence space with $\nu$\{$p_{3/2}$, $p_{1/2}$, $s_{1/2}$, $d_{5/2}$\} using the Berggren basis and remaining higher partial waves using the HO basis. 

As benchmarking for the EC calculation, we first test $^4$H and $^6$He nuclei. $^4$H has a $2^-$ ground state (g.s.) which is a broad resonance~\cite{PhysRevC.100.054313}, while $^6$He becomes a resonance in excited states~\cite{TILLEY20023}. In the $^4$H calculation, we use the chiral N$^3$LO NN interaction developed by Entem and Machleidt (labeled by $\mathrm{N^3LO_{EM}}$)~\cite{PhysRevC.68.041001}, and then evolve it to a low-momentum scale $\lambda=2.0$ $\mathrm{fm}^{-1}$ using the similarity renormalization group (SRG)~\cite{PhysRevC.75.061001,BOGNER201094}. In the $^6$He calculation based on GSM with the $^4$He core, we use the COSM coordinates to reduce the COM effect, as mentioned already. In the COSM coordinates, however, the effective interaction between valence particles cannot yet be obtained from realistic nuclear forces~\cite{Michel_2009, SUN2017227}. Therefore, we use the FHT potential~\cite{PhysRevC.96.054316} for the valence-particle effective interaction in the GSM calculation with a core. For $^4$H and $^6$He, Hamiltonians at each $\alpha$ point (even at the target point $\alpha_*$) can be solved directly by NCGSM and GSM with $^4$He core, respectively. As a benchmark for EC emulations, we refer to the NCGSM (GSM) calculation as the ``exact'' solution.

\begin{figure*}[t]
    \centering
    \includegraphics[width=0.75\textwidth]{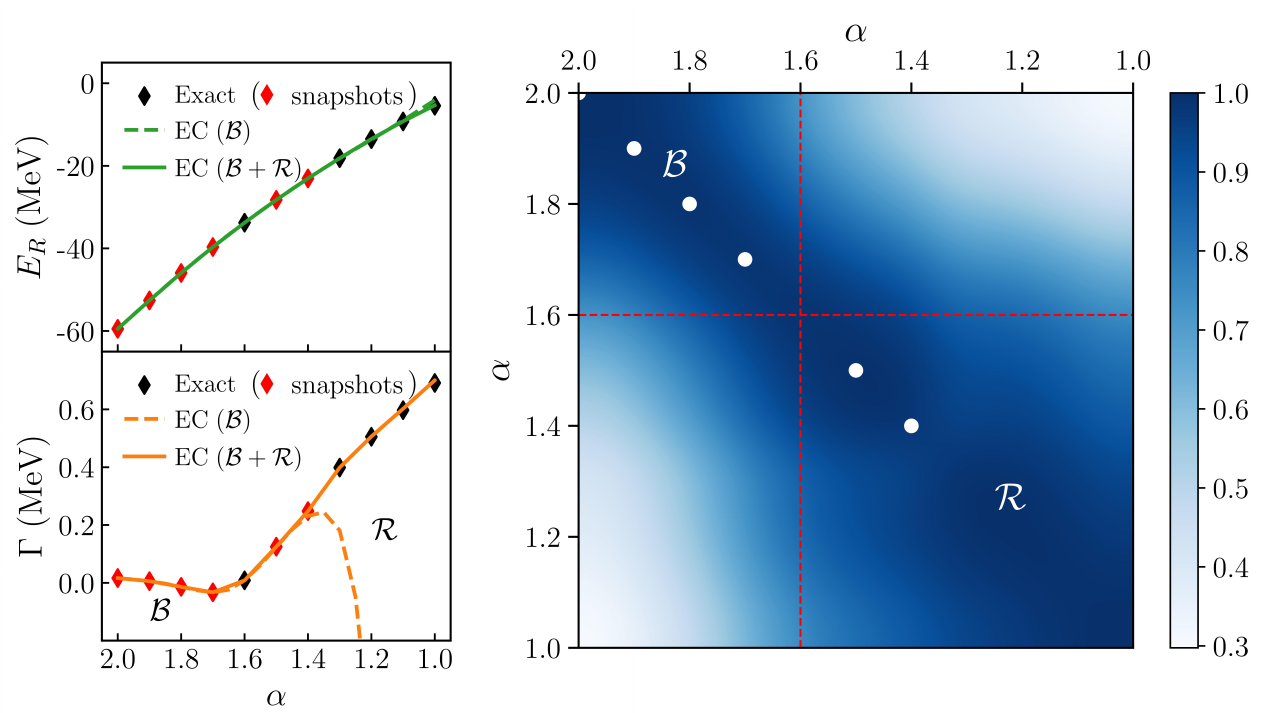}
    \caption{$^4$H EC extrapolation to the $2^{-}$ g.s. broad resonance based on NCGSM with $\mathrm{{N^3LO}_{EM}}$. Left panels show the real part $E_R$ and resonance width $\Gamma$ of the eigen energy obtained using the EC emulator based on $\alpha=1.7, 1.8, 1.9, 2.0$ bound snapshots [denoted by EC($\mathcal{B}$)], and using the emulator with $\alpha=1.4,1.5$ narrow resonances incorporated [denoted by EC($\mathcal{B+R}$)], compared with exact NCGSM calculations. The right panel shows norm matrix elements $\tilde{N}_{ij}$ in the grid of the control parameter $\alpha$ with the chosen snapshots indicated by white dots. The bound and resonance regimes are indicated by $\mathcal{B}$ and $\mathcal{R}$, respectively. Note that norm matrix elements are complex numbers, therefore we plot their absolute values.}
    \label{fig:4H}
\end{figure*}

In the EC calculation by scaling the interaction $V^{\mathrm{NN}}$ with a factor of $\alpha>1$, the system with the Hamiltonian $H(\alpha)$ can evolve into a narrow resonance and eventually a bound state. For $^4$H, $\alpha\gtrsim 1.6$ can lead to a bound ground state. Therefore, we take $\alpha=1.7$, 1.8, 1.9 and 2.0 as EC snapshots which belong to the bound regime labeled by $\mathcal{B}$. For $\alpha<1.6$, the Hamiltonian $H(\alpha)$ gives a resonance $^4$H. We take $\alpha=1.4$ and 1.5 as snapshots which belong to the resonance regime labeled by $\mathcal{R}$. We have tested the EC emulator constructed by only $\alpha=1.7$, 1.8, 1.9 and 2.0 bound snapshots, denoted by EC($\mathcal{B}$). Using this emulator, eigenvalues at other $\alpha$ points are obtained through the EC emulation. As shown in the top left panel of Fig.~\ref{fig:4H}, even using only the bound snapshots, the real part of the eigen energy of a resonance beyond the training regime agrees well with the exact solution obtained by the NCGSM. However, the imaginary part (resonance width $\Gamma$) of the eigen energy of a broad resonance at $\alpha<1.4$ deviates severely from the exact solution, as shown in the lower left panel of Fig.~\ref{fig:4H}. But, if we incorporate $\alpha=1.4$ and 1.5 narrow resonances into the EC snapshots [denoted by EC($\mathcal{B+R}$)], both energy and resonance width can be well obtained (including at the $\alpha_*$ target point) compared with exact solutions given by NCGSM. This means that a good EC emulator for extrapolation to the regime of broad resonances should include snapshots from narrow resonances. In Refs.~\cite{Yapa2023,Yapa2025}, the authors used the complex conjugates of complex-scaling-rotated training bound states to provide resonance ingredients for the EC extrapolation to the target resonance.

In the right panel of Fig.~\ref{fig:4H}, we show norm matrix elements calculated by Eq.~(\ref{norm}). It is seen that, when $|\psi(\alpha_i)\rangle$ and $|\psi(\alpha_j)\rangle$ belong to the same regime, the value of the norm matrix element is close to 1, indicating that the wave functions have similar structures. When $|\psi(\alpha_i)\rangle$ and $|\psi(\alpha_j)\rangle$ come from different regimes, the value of the norm matrix element is significantly reduced, indicating different structures of the wave functions. This suggests that one should add the components of narrow resonances to the EC emulator for extrapolation to broad resonances. It was commented that both complex conjugates of rotated training bound states and narrow resonances can be used to provide EC snapshots for extrapolation to broad resonances~\cite{Yapa2025}.

\begin{figure}[t]
    \includegraphics[width=0.482\textwidth]{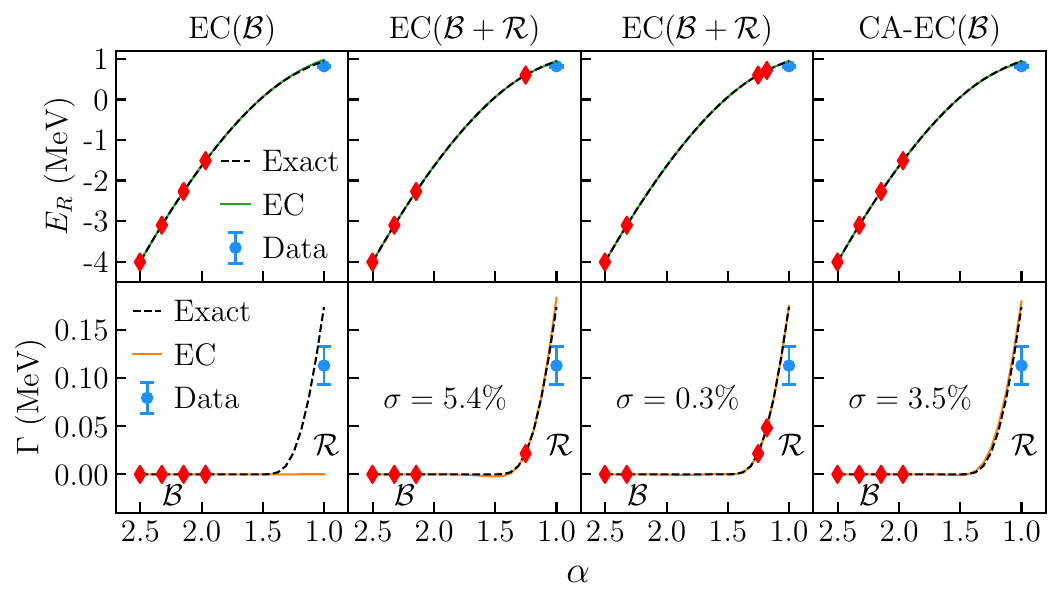}
    \centering
    \caption{EC emulated energy $E_R$ (relative to $^4$He) and resonance width $\Gamma$ of the $^6$He $2^+_1$ excited state using three different choices of four snapshots, compared with exact GSM calculations and experimental data~\cite{6He_exp}. The deviation from the exact width at the target point $\alpha_*=1.0$ is defined by 
    $\sigma=|\Gamma^{\mathrm{EC}}-\Gamma^{\mathrm{Exact}}|/\Gamma^{\mathrm{Exact}}$. EC calculation with the choice of only the four bound snapshots leads to a bound extrapolation result, as shown in the leftmost panel.}
    \label{fig:6He}
\end{figure}

$^6$He is a weakly bound nucleus in the $0^+$ ground state, but becomes resonance in excited states. The first excited state is a $2^+$ resonance~\cite{TILLEY20023}. We adopt the valence-space COSM Hamiltonian with the FHT potential which was used in the previous GSM calculations~\cite{PhysRevC.104.L061301,PhysRevC.96.054316,PhysRevC.102.024309,9Nexp2023}. A scaling factor of $\alpha\gtrsim 1.3$ can turn the $2^+_1$ resonance into a bound state. To illustrate the detailed sensitivity of the EC emulator to the location of snapshots, we choose different sets of snapshots, and test the qualities of the emulators across the whole regime from $\alpha=2.5$ to $1.0$. Figure~\ref{fig:6He} shows EC emulations with three different emulators based on three different choices of four snapshots, compared with exact solutions obtained by the GSM with the $^4$He core. As seen in Fig.~\ref{fig:6He}, if we choose four snapshots all from the bound regime $\mathcal{B}$, the EC emulator cannot lead to any resonance solutions. It may be naturally understood since this emulator does not contain any resonance ingredients. If we choose three (or two) snapshots from the bound regime $\mathcal{B}$ and one (or two) from the narrow resonance regime $\mathcal{R}$, the emulator can well reproduce exact solutions in the whole regime from $\alpha=2.5$ to 1.0. In Refs.~\cite{Yapa2023,Yapa2025}, the use of complex conjugates of rotated bound snapshots provides resonance components. Figure~\ref{fig:6He} also includes complex-augmented eigenvector continuation (CA-EC) calculations~\cite{Yapa2023,Yapa2025} using the four bound snapshots and their complex conjugates. We see that the EC extrapolation with two narrow resonance snapshots used increases the accuracy of calculation giving a smaller deviation as shown in Fig.~\ref{fig:6He}. However, narrow resonance snapshots should be harder to be calculated than bound snapshots, and the CA-EC method is particularly useful when no direct resonance solution is feasible. From the calculations, we see that narrow resonance snapshot(s) or bound complex conjugates should be required in EC extrapolations to broad resonances.

$^4$H and $^6$He resonances have provided excellent testing grounds for EC calculations, in which direct NCGSM or GSM calculations can be used as benchmarks. In addition, $^4$H and $^6$He are also of particular interests in the field of exotic nuclei. Due to the character of unbound broad resonance, $^4$H is difficult to be produced experimentally, and experiments so far have not given consistent data with energy ranging from 1.6 to 3.8 MeV (relative to the $^3{\text H}+n$ threshold) and resonance width ranging from 0.4 to 4.7 MeV~\cite{PhysRevC.104.024319}. Theoretical calculations have also given different results. NCGSM with realistic interactions gave the $^4$H energy around 1.7 MeV (above the $^3{\text H}+n$ threshold) and width around 0.9 MeV~\cite{PhysRevC.104.024319}, while NCGSM with phenomenological interactions gave the similar energy but larger width by a factor 2~\cite{PhysRevC.104.L061306}. The Faddeev-Yakubovsky equations gave an energy about 1.2 MeV but a large width about 4 MeV~\cite{LAZAUSKAS2019335}. The present calculation gives an energy of 2.65 MeV and a width of 0.7 MeV. $^6$He is the lightest Borromean halo~\cite{PhysRevLett.93.142501,PhysRevC.84.051304}. It provides us with a concise picture as an open quantum system with two weakly-bound or unbound resonant valence neutrons outside the deeply bound $^4$He core, and hence provides a good testing ground for nucleon-nucleon interactions and continuum effects. The present calculation gives an energy of 0.947 MeV (relative to $^4$He) and width of 174 keV for the $2_1^+$ resonance, consistent with previous calculations with energy in a range from 0.851~\cite{PhysRevC.84.051304} to 1.01 MeV~\cite{PhysRevC.98.061302} and width in a range from 109~\cite{PhysRevC.84.051304} to 207 keV~\cite{PhysRevC.98.061302}. These results are in reasonable agreements with the experimental energy of 0.824 MeV and width of 113 keV~\cite{6He_exp}.

${}^{4}n$ and $^7$He systems are typical broad resonances for which direct \textit{ab initio} calculations have not been achieved, therefore extrapolations have to be used~\cite{PhysRevLett.117.182502,PhysRevLett.119.032501,PhysRevC.100.054313,PhysRevLett.118.232501}. Usually, polynomial extrapolations were used to reach the target state~\cite{PhysRevLett.117.182502,PhysRevLett.119.032501,PhysRevC.100.054313,PhysRevLett.118.232501}, which requires actually monotonic variations of the energy and width with the control parameter. In Refs.~\cite{Frame2018,PhysRevLett.123.252501,PhysRevC.101.041302,KONIG2020135814,Yapa2023,Yapa2025}, it has been commented that the EC extrapolation should be more reliable. 

\begin{figure}[t]
    \includegraphics[width=0.4\textwidth]{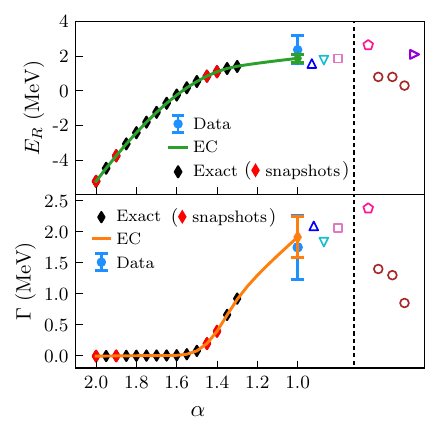}
    \centering
    \caption{EC calculations of ${}^{4}n$ energy $E_\textsc{R}$ and resonance width $\Gamma$, compared with the data of the observed resonance-like peak~\cite{Duer2022}. For $\mathrm{N^3LO_{EM}}$, we present detailed EC result as a function of $\alpha$, with theoretical uncertainties at the target $\alpha_*=1.0$ indicated by error bars. EC extrapolations at the target with $\mathrm{N^2LO_{opt}}$, $\mathrm{N^4LO_{EMN}}$ and $\mathrm{N^3LO_{local}}$ are shown by symbols {\color{blue}\scalebox{1.2}{$\bm{\vartriangle}$}}, {\color{C9}\scalebox{1.2}{$\bm{\triangledown}$}} and {\color{C6}\scalebox{1.0}{$\bm{\amssBox}$}}, respectively, in the right. Other calculations based on NCGSM ({\color{DeepPink}\scalebox{1.17}{$\pmb{\wasyPentagon}$}})~\cite{PhysRevC.100.054313}, NCSM ({\color{brown}\scalebox{1.8}{$\raisebox{-0.35ex}{$\bm{\circ}$}$}})~\cite{PhysRevLett.117.182502} and QMC ({\color{violet}\scalebox{1.8}{$\raisebox{-0.3ex}{${\triangleright}$}$}})~\cite{PhysRevLett.118.232501} are also included in the rightmost for comparisons. }
    \label{fig:4n}
\end{figure}

In the EC calculation of the ${}^{4}n$ system, to test the sensitivity to the interaction used, we have also used other three chiral forces, $\mathrm{N^2LO_{opt}}$~\cite{PhysRevLett.110.192502}, 
$\mathrm{N^4LO_{EMN}}$~\cite{PhysRevC.96.024004}
and $\mathrm{N^3LO_{local}}$~\cite{PhysRevC.107.034002}, in addition to the SRG-evolved $\lambda=2.0$ $\mathrm{fm}^{-1}$ $\mathrm{N^3LO_{EM}}$~\cite{PhysRevC.68.041001} interaction which was used in the $^4$H calculation. $\mathrm{N^4LO_{EMN}}$ and $\mathrm{N^3LO_{local}}$ interactions were evolved to a low-momentum scale $\lambda=2.0$ $\mathrm{fm}^{-1}$ using SRG as $\mathrm{N^3LO_{EM}}$. Figure~\ref{fig:4n} shows the EC calculations of the ${}^{4}n$ energy and resonance width. The EC emulator was built on four snapshots (two bound and two narrow resonance points). EC results are not sensitive to the locations of chosen bound and narrow resonance snapshots. Exact NCGSM calculations are converged in the regime of $\alpha\gtrsim 1.3$, while the result at the target point $\alpha_*=1.0$ is obtained using EC extrapolation. We see that the four chiral interactions lead to similar EC extrapolated energies and resonance widths of the ${}^{4}n$ system. The results are close to the data of the ${}^{4}n$ resonance-like peak observed in the missing mass spectrum~\cite{Duer2022}. For comparisons, Fig.~\ref{fig:4n} also plots previous calculations with polynomial extrapolations based on NCGSM~\cite{PhysRevC.100.054313}, no-core shell model (NCSM)~\cite{PhysRevLett.117.182502} and quantum Monte Carlo (QMC)~\cite{PhysRevLett.118.232501}.

The ${}^{4}n$ system has a long history of search both theoretically and experimentally. Most theoretical calculations predict a ${}^{4}n$ resonance, e.g., in Refs.~\cite{PhysRevLett.90.252501,PhysRevLett.117.182502,PhysRevLett.118.232501,PhysRevLett.119.032501,PhysRevC.100.054313}. However, the peak observed in the ${}^{4}n$ missing mass spectrum~\cite{Duer2022} has also been explained as a consequence of dineutron-dineutron correlations by reaction calculations~\cite{PhysRevLett.130.102501}. Experimentally, ${}^{4}n$ direct detection has not been accessible. The experiment~\cite{Duer2022} used the knockout reaction $^8$He($p$, $p^4$He) to produce the ${}^{4}n$ system. The ${}^{4}n$ missing mass spectrum was obtained by the mass reconstruction with the knocked-out $\alpha$ particle and scattered proton~\cite{Duer2022}. The observed peak was suggested to have a ${}^{4}n$ resonance-like structure~\cite{Duer2022}. A future ${}^{4}n$ direct detection would be of most value~\cite{NatureNews}.

The unbound nucleus ${}^{7}$He provides another challenging case that demonstrates the power of EC extrapolation in describing broad resonances. From experimental data~\cite{TILLEY20023}, the $3/2^-$ g.s. is a narrow resonance, while the second and third excited states with $J^\pi=1/2^-$ and $5/2^-$, respectively, are broad resonances. Exact GSM calculation with the $^4$He core can lead to a converged result for the $3/2^-$ g.s. narrow resonance, but it is not converged for the $1/2^-$ and $5/2^-$ excited broad resonances due to computational instability during the diagonalizing of the complex Hamiltonian. This difficulty originates from the strong coupling to the continuum in the broad resonances, and could only be partly overcome using the density matrix renormalization group approach as in the previous study~\cite{PhysRevLett.97.110603}.

\begin{figure}[t]
    \includegraphics[width=0.482\textwidth]{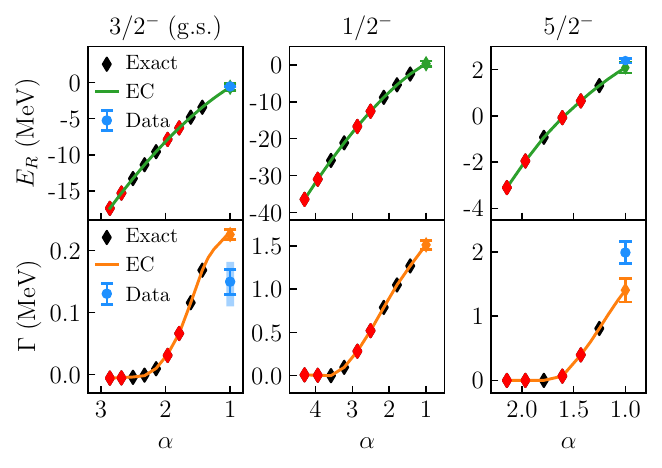}
    \centering
    \caption{EC emulations for ${}^{7}$He $3/2^-$ g.s., $1/2^-$ and $5/2^-$ excited states, with theoretical uncertainties at the target $\alpha_*=1.0$ indicated by error bars. Red diamonds indicate chosen EC snapshots which belong to exact GSM solutions with $^4$He core. Experimental data are taken from Ref.~\cite{TILLEY20023}, but for the $3/2^-$ resonance different experiments~\cite{TILLEY20023,CAO201246,PhysRevC.109.L061602} gave different widths indicated by a blue shadowing bar. Note that different axis scales are used for different panels.}
\label{fig:7He}
\end{figure}

Figure~\ref{fig:7He} shows EC extrapolations for the $3/2^-$ g.s., $1/2^-$ and $5/2^-$ excited states in ${}^{7}$He using the same interaction and technique as in the ${}^{6}$He calculation. We choose two bound and two resonance snapshots (indicated by red diamonds in Fig.~\ref{fig:7He}) for which exact GSM calculations with the $^4$He core are converged. We see that EC emulations agree well with exact GSM calculations. The EC extrapolated energies agree well with the data~\cite{TILLEY20023}, but the extrapolated width is larger than data in the $3/2^-$ g.s., and smaller than data in the $5/2^-$ excited state. For the $1/2^-$ excited state, existing experimental information is not firm~\cite{PhysRevC.72.061301,PhysRevLett.95.132502,PhysRevLett.88.102501} (so that we are unable to plot the conflicting data in Fig.~\ref{fig:7He}). As shown in Table~\ref{table_7He}, theoretical investigations have also given very different results.

\begin{table}[hbtp]
    \centering
    \caption{Calculated energies $E_R$ (relative to $^6\mathrm{He}+n$ threshold, in MeV) and widths $\Gamma$ (in MeV) of the $^7$He resonances, compared with other theoretical calculations collected in~\cite{PhysRevC.106.064320} and data~\cite{TILLEY20023,CAO201246,PhysRevC.109.L061602,PhysRevC.72.061301,PhysRevLett.95.132502,PhysRevLett.88.102501}. Different theories (experiments) might give significantly different results, which are indicated by ranges.}
    \renewcommand{\arraystretch}{1.35}
    \setlength{\tabcolsep}{0.9mm}
    \begin{tabular}{ccccccc}
    \hline\hline
    & \multicolumn{2}{c}{This work} & \multicolumn{2}{c}{Other cal.} & \multicolumn{2}{c}{Expt.} \\
    $J^\pi$ & $E_R$ & $\Gamma$ & $E_R$ & $\Gamma$ & $E_R$ & $\Gamma$ \\
    \hline
    $\frac{3}{2}^-$ & 0.50 & 0.23 & 0.28$-$0.71 & 0.05$-$0.57 & 0.430 & 0.11$-$0.182\\
    $\frac{1}{2}^-$ & 1.48 & 1.51 & 1.81$-$2.70 & 2.07$-$5.0 & 1.0$-$3.5 & 0.75$-$10\\
    $\frac{5}{2}^-$ & 3.23 & 1.40 & 3.13$-$4.10 & 1.07$-$2.30 & 3.36(9) & 1.99(17)\\
    \hline\hline
\end{tabular}
\label{table_7He}
\end{table}

We have also estimated theoretical uncertainties that mainly come from: i) NCGSM (GSM) model space and basis potential parameters; ii) EC extrapolation; iii) NN interactions and SRG evolution if the EFT force is used. These uncertainties have been analyzed term by term as follows:
\begin{enumerate}[label=(\alph*)]
\item In the previous works~\cite{PhysRevLett.119.032501,PhysRevC.100.054313,PhysRevC.96.054316}, uncertainties from NCGSM (GSM) model space and basis parameters had been well analyzed. For example, it had been tested that effects on the $^4n$ calculation by changing basis parameters by 15\% and removing the $g_{7/2}$ high partial wave are tiny~\cite{PhysRevLett.119.032501}. In the present work, we have checked that the above changes in basis parameters and model space bring about 50 keV uncertainty for both energy and width in $^4n$ and $^4$H calculations. This uncertainty is even smaller for $^6$He and $^7$He GSM calculations. The use of natural orbitals has improved convergences.
\item The similar technique as in~\cite{Yapa2023} has been used to estimate the uncertainty of the EC extrapolation by randomly choosing two bound and two resonance snapshots, and the resulted uncertainty was quantified by one standard deviation around the median value, giving that EC uncertainty is about 60 keV for both energy and width in $^4n$ and $^4$H. This uncertainty is about 10 keV in the EC extrapolations of $^6$He and $^7$He. We have also found that taking more snapshots do not significantly improve the extrapolation result. This shows that the EC($\mathcal{B+R}$) extrapolation uncertainty in present calculations is well controlled.
\item We find that the dominate uncertainty is from interactions used. Although a more rigorous EFT uncertainty analysis is possible, we have tried to use different sets of nuclear forces at different chiral orders (from N$^2$LO to N$^4$LO) to justify that the $^4n$ extrapolation based on NCGSM calculations is not very sensitive to the underlying nuclear force, particularly considering large experimental errors. To be specific, by using different chiral nuclear forces, the calculated $^4n$ energy and width vary 0.3 MeV and 0.5 MeV at most, respectively. Three-nucleon force (3NF) is not included because its inclusion greatly increases the complexity of NCGSM calculations where resonance and continuum partial waves are involved. In $^4n$, only the weak $T=3/2$ channel of 3NF appears. Though the dominate $T=1/2$ channel of 3NF exists in $^4$H, SRG-induced and genuine 3NFs almost cancel each other out in $A=4$ systems when the SRG flow parameter is in the range of $\lambda\approx 1.6$$-$2.0~fm$^{-1}$, and calculated energy is insensitive to $\lambda$~\cite{PhysRevLett.107.072501}. This meanwhile indicates the uncertainty from the SRG evolution is small if $\lambda\approx 1.6$$-$2.0~fm$^{-1}$is used in $A=4$ calculations. We have tested that the $^4n$ ($^4$H) energy and width change about 0.1 MeV and 0.2 MeV, respectively, when $\lambda$ is changed from 2.0 to 1.8~fm$^{-1}$. In Fig.~\ref{fig:4n}, we plot the estimated total uncertainties of the target $^4n$ energy and width.  
For $^6$He and $^7$He, the dominate theoretical uncertainty is also from the interaction used. We have estimated that uncertainties corresponding to statistical uncertainties of FHT parameters~\cite{PhysRevC.96.054316,PhysRevC.102.024309} in $^7$He are: 0.49 MeV (in energy) and 8 keV (in width) for the $3/2^-$ g.s.; 0.75 MeV and 48 keV, respectively, for the $1/2^-$ energy and width; and 0.24 MeV and 184 keV, respectively, for the $5/2^-$ energy and width, plotted in Fig.~\ref{fig:7He}.
\end{enumerate}

\section{Summary}
How to treat broad resonances of nuclei is challenging current theories and computations. In the present work, we have applied the eigenvector continuation method to broad resonances of real nuclear systems. The extension of the stationary Schr{\"o}dinger solution to the complex-energy domain can well treat the coupling to the continuum which is an essential component in resonance states. However, the inclusion of the continuum coupling drastically increases the dimension of the model, and therefore poses computational challenges, making numerical calculations non-convergent or unstable. By scaling the interaction (or part of the interaction), the eigenvector continuation uses a few artificial bound and narrow resonance states to extrapolate the broad resonance of the target Hamiltonian. This is a diagonalizing process of the target Hamiltonian within a subspace constructed by the converged training points of the scaled Hamiltonian, rather than a simple monotonic extrapolation. 

In the present work, we have well tested the complex-energy eigenvector continuation by investigating $^4$H, four-neutron $^4n$, $^6$He and $^7$He resonances. The no-core Gamow shell model (or with core) can provide converged bound and narrow resonance solutions with scaled nuclear forces, while the broad resonance solution is obtained through the eigenvector continuation. Eigenvector continuation solutions have been well benchmarked against exact no-core Gamow shell model (or with core) calculations in $^4$H and $^6$He. Resonance energies and widths obtained by the eigenvector continuation agree with available experimental data. The experimental energy and width of the $1/2^-$ broad resonance in $^7$He remain rather uncertain. The present predictions of the complex-energy eigenvector continuation should be useful for future experiments.

\section{Acknowledgements}
We thank R. Machleidt for providing us with the local position-space NN potential. This work has been supported by the National Key R\&D Program of China under Grants No. 2024YFA1610900, and No. 2023YFA1606401; the National Natural Science Foundation of China under Grants No. 12335007, No. 12535008, No. 12035001, No. 12205340, No. 12347106 and No. 12121005. We acknowledge the High-Performance Computing Platform of Peking University for providing computational resources.

\bibliographystyle{elsarticle}
\bibliography{reference}
\end{document}